\documentclass [12pt]{article}

\begin{document}
\begin{center}
\textbf{Relationship between Conformal Geometrodynamics and Dirac Equations 
}
\end{center}

\begin{center}
\textbf{M.V.Gorbatenko}
\end{center}

\begin{center}
Russian Federal Nuclear Center -- All-Russian Research Institute of 
Experimental Physics, Sarov, N.Novgorod region; E-mail: \underline {gorbatenko@vniief.ru}
\end{center}

\begin{center}
\textbf{Abstract}
\end{center}

The paper describes a unique phenomenon -- the possibility of establishing, 
in certain space regions, the one-to-one correspondence between equations 
related to absolutely different physical phenomena: (1) phenomena associated 
with the Weyl degrees of freedom in plane space; (2) phenomena, which can be 
described in terms of half-integer spin particles and observed quantities 
corresponding to a full set of bispinors. 

The phenomenon established opens wide prospects for resolving in future the 
``old'' disputable issue concerning the physical meaning of the Weyl vector. 
The paper discusses, in particular, the possibility of identifying the Weyl 
vector with the current density vector of bispinors constituting a bispinor 
matrix included in the Dirac equation. Some other issues are also discussed.

\section{Introduction}
\label{sec:introductionon}

To date, there are two approaches that provided results, which are of 
interest as applied to the development of a unified theory of space and 
three fundamental interactions -- strong, electromagnetic, and weak. 
Conformal geometrodynamics (CGD)\footnote{ For explanations concerning CGD 
equations (including their history) see \cite{Gorb2005}, \cite{GorbRom2005} 
and the papers they refer to.} is one these 
approaches. This term means the theory based on the conformally invariant 
generalization of the general relativity (GR) equations. Another approach is 
studying the algorithm of constructing a bispinor system representing the 
given system of tensor fields. The algorithm has been developed and 
described in \cite{GorbPush2001} - \cite{GorbMono}. 

In \cite{Gorb2005-2} the authors made an attempt to use the 
algorithm above for correlating the new degrees of freedom in CGD equations 
and the bispinors following the Dirac equation. The given paper proposes, in 
essence, an improved way (in comparison with \cite{Gorb2005-2}) of setting up the 
correlation and interpreting the physical meaning of the Weyl vector 
appeared in CGD equations. 

The improvement consists, first of all, in ascertaining, what particular CGD 
equations we are speaking about. We operate equations following from 
equation $T_\varepsilon {}^\mu _{;\mu } = 0$, where $T_{\alpha \beta } $ is the 
energy-momentum tensor constructed from the Weyl degrees of freedom and 
included in the right-hand member of the conformally generalized GR 
equations. Space is considered to be a plane, the inverse impact of the Weyl 
vector field on the space curvature is ignored. Though we deal with plane 
space, the equations of interest describe the physics of the Weyl degrees of 
freedom of a general kind; the Weyl vector is not reduced to a scalar 
function gradient in our approach.

The second improvement is that incompleteness (in \cite{Gorb2005-2}) of the obtained set 
of equations for the observed tensors, which follow from the Dirac equation, 
has been avoided. It is assumed that the Dirac equation is related to a set 
of four bispinors reproducing the given tensor values, rather than the one 
bispinor. Such an approach follows the algorithm of mapping tensors onto 
bispinors published earlier in \cite{GorbPush2001} - \cite{GorbMono}. 
A new result of 
this study includes expressions for the matrix connectivity correlated with 
bispinor matrix equations.

These improvements allow us to assert that in certain spacetime regions, 
where the polarization matrix is positively defined, solutions of CGD 
equations can be interpreted in terms of solutions of the Dirac equation. 
This interpretation automatically leads to equations defining the matrix 
connectivity (in other terms, gauge field) and symmetry groups. Proving 
these facts constitutes the major result of this study. The result, which 
confirms the interpretation of the Weyl vector in \cite{Gorb2005-2} 
as a current density vector for the whole set 
of bispinors making up the bispinor matrix in the Dirac equation. 

The results obtained are summarized at the end of the paper. We also discuss 
prospects of using the results of this study as a basis for a new approach 
to setting up the unified theory of gravitational and Weyl degrees of 
freedom and internal degrees of freedom, which are required for describing 
the whole diversity of half-integer spin particles. 

\section{Dirac Equation and Its Corollaries}
\label{sec:dirac}

\subsection{Dirac Matrices. Dirac Equation}
\label{subsec:dirac}

The Dirac matrices (DMs) $\tilde {\gamma }_\alpha $ are defined as

\begin{equation}
\label{eq1}
\tilde {\gamma }_\alpha \tilde {\gamma }_\beta + \tilde {\gamma }_\beta 
\tilde {\gamma }_\alpha = 2\eta _{\alpha \beta } .
\end{equation}

They are constant all over the space. If an explicit form of DMs is 
required, we use the Majoran set of matrices

\begin{equation}
\label{eq2}
\tilde {\gamma }_0 = - i\rho _2 \sigma _1 ,\quad \tilde {\gamma }_1 = \rho 
_1 ,\quad \tilde {\gamma }_2 = \rho _2 \sigma _2 ,\quad \tilde {\gamma }_3 = 
\rho _3 ,
\end{equation}

\noindent
the elements of which are integer real numbers\footnote{ Solutions of 
relationships (\ref{eq1}) always include a real set of DMs, if the signature $\left( 
{ - + + + } \right)$ is used. }. 

Suppose that the field operator $Z$ in the general case is a $4\times 4_{ 
}$matrix satisfying the Dirac equation

\begin{equation}
\label{eq3}
\tilde {\gamma }^\nu \left( {\nabla _\nu Z} \right) = m \cdot Z.
\end{equation}

The matrix operator $Z_{ }$will be called a bispinor matrix. Bispinor 
states are separated from $Z_{ }$by its multiplication on the right by 
projection operators.

Together with Eq. (\ref{eq3}), the following equation is satisfied:

\begin{equation}
\label{eq4}
\left( {\nabla _\nu Z^ + } \right)\tilde {\gamma }^\nu \tilde {D} = - m 
\cdot Z^ + \tilde {D}.
\end{equation}

The matrix $\tilde {D}_{ }$found in (\ref{eq4}) is defined as

\begin{equation}
\label{eq5}
\tilde {D}\tilde {\gamma }_\mu \tilde {D}^{ - 1} = - \tilde {\gamma }_\mu ^ 
+ .
\end{equation}

The covariant derivatives of the bispinor matrix in (\ref{eq3}), (\ref{eq4}) are written as

\begin{equation}
\label{eq6}
\left. {\begin{array}{l}
 \nabla _\alpha Z = Z_{;\alpha } - Z\Gamma _\alpha \\ 
 \nabla _\alpha Z^ + = Z^ + _{;\alpha } + \Gamma _\alpha Z^+ \\ 
 \end{array}} \right\} .
\end{equation}

The quantity $\Gamma _\alpha $ in (\ref{eq6}) will denote matrix connectivity. It is 
a set of real skew-Hermitian matrices:

\begin{equation}
\label{eq7}
\Gamma ^\ast _\alpha = \Gamma _\alpha ,\quad \Gamma ^ + _\alpha = - \Gamma 
_\alpha .
\end{equation}

The matrix skew $P_{\alpha \beta } $ is defined through the matrix 
connectivity $\Gamma _\alpha $ in much the same way as the Riemann tensor is 
defined through the Christoffel symbols.

\begin{equation}
\label{eq8}
P_{\alpha \beta } = \Gamma _{\beta ,\alpha } - \Gamma _{\alpha ,\beta } + 
\Gamma _\alpha \Gamma _\beta - \Gamma _\beta \Gamma _\alpha .
\end{equation}

The following equalities hold:

\begin{equation}
\label{eq9}
\left. {\begin{array}{l}
 \left( {\nabla _\alpha \nabla _\beta - \nabla _\beta \nabla _\alpha } 
\right)Z = - ZP_{\alpha \beta } \\ 
 \left( {\nabla _\alpha \nabla _\beta - \nabla _\beta \nabla _\alpha } 
\right)Z^ + = P_{\alpha \beta } Z^ + \\ 
 \end{array}} \right\} .
\end{equation}

Let us give some relationships that follow immediately from the Dirac 
equation. Eq. (\ref{eq3}) on the left is multiplied by $\tilde {\gamma }^\alpha $.

\begin{equation}
\label{eq10}
\tilde {\gamma }^\alpha \tilde {\gamma }^\nu \left( {\nabla _\nu Z} \right) 
= m \cdot \tilde {\gamma }^\alpha Z.
\end{equation}

The product $\tilde {\gamma }^\alpha \tilde {\gamma }^\nu $ is written as

\begin{equation}
\label{eq11}
\tilde {\gamma }^\alpha \tilde {\gamma }^\nu = \eta ^{\alpha \nu } + \tilde 
{S}^{\alpha \nu }
\end{equation}

(here, $\tilde {S}_{\mu \nu } = \frac{1}{2}\left( {\tilde {\gamma }_\mu 
\tilde {\gamma }_\nu - \tilde {\gamma }_\nu \tilde {\gamma }_\mu } \right))$ 
and (\ref{eq11}) is substituted into (\ref{eq10}).

\begin{equation}
\label{eq12}
\left( {\nabla _\alpha Z} \right) = - \tilde {S}_\alpha ^\nu \left( {\nabla 
_\nu Z} \right) + m \cdot \tilde {\gamma }_\alpha Z.
\end{equation}

After Hermitian conjugation of Eq. (\ref{eq12}) and multiplication by$D$, we obtain:

\begin{equation}
\label{eq13}
\left( {\nabla _\alpha Z^ + } \right)\tilde {D} = \left( {\nabla _\nu Z^ + } 
\right)\tilde {D}\tilde {S}_\alpha ^\nu - m \cdot Z^ + \tilde {D}\tilde 
{\gamma }_\alpha .
\end{equation}

Then, Eq. (\ref{eq3}) on the left is multiplied by $\tilde {\gamma }^\mu \nabla _\mu 
$.

\begin{equation}
\label{eq14}
\tilde {\gamma }^\mu \nabla _\mu \tilde {\gamma }^\nu \left( {\nabla _\nu Z} 
\right) = m \cdot \tilde {\gamma }^\mu \nabla _\mu Z.
\end{equation}

Note that within the formalism at issue

\begin{equation}
\label{eq15}
\nabla _\mu \tilde {\gamma }^\alpha = 0.
\end{equation}

As a result, we have:

\begin{equation}
\label{eq16}
\left( {\eta ^{\mu \nu } + \tilde {S}^{\mu \nu }} \right)\left( {\nabla _\mu 
\nabla _\nu Z} \right) = m^2 \cdot Z.
\end{equation}

It follows from Eq.(\ref{eq16}) that

\begin{equation}
\label{eq17}
\left. {\begin{array}{l}
 \left( {\nabla ^\nu \nabla _\nu Z} \right) = \frac{1}{2}\tilde {S}^{\mu \nu 
}ZP_{\mu \nu } + m^2 \cdot Z \\ 
 \left( {\nabla ^\nu \nabla _\nu Z^ + } \right)\tilde {D} = 
\frac{1}{2}ZP_{\mu \nu } Z^ + \tilde {D}\tilde {S}^{\mu \nu } + m^2 \cdot Z^ 
+ \tilde {D} \\ 
 \end{array}} \right\} .
\end{equation}

The second relationship in Eq. (\ref{eq17}) is derived from the first one by the 
Hermitian conjugation.

\subsection{Algorithm for Tensor Mapping onto Bispinor Matrix}
\label{subsec:algorithm}

We define the vector $j_\alpha $ and the anti-symmetric tensor $h_{\alpha 
\beta } $ as follows:

\begin{equation}
\label{eq18}
j_\alpha \equiv \frac{1}{4}\mbox{Sp}\left\{ {Z^ + \tilde {D}\tilde {\gamma 
}_\alpha Z} \right\},
\end{equation}

\begin{equation}
\label{eq19}
h_{\alpha \beta } \equiv - \frac{1}{8}\mbox{Sp}\left\{ {Z^ + \tilde 
{D}\tilde {S}_{\alpha \beta } Z} \right\}.
\end{equation}

Formulas (\ref{eq18}), (\ref{eq19}) make it possible to uniquely define the vector $j_\alpha 
$ and the anti-symmetric tensor $h_{\alpha \beta } $ based on the 
real bispinor matrix $Z$. 

As noted above, in Refs. \cite{GorbPush2001} - \cite{GorbMono}, 
an inverse task 
was accomplished , i.e. the task of finding the bispinor matrix $Z_{ 
}$based on the given vector $j_\alpha $ and antisymmetric tensor 
$h_{\alpha \beta } $. Refs. 
[3]-[5] addressed a complex 
case, in which the sought bispinor matrix $Z$ reproduced the whole given set 
of tensor quantities (scalar, pseudoscalar, vector, pseudovector, 
antisymmetric tensor). This paper deals with a particular case of DM and 
matrix $Z$ implementations over the field of real numbers. In this case, 
only the vector $j_\alpha $ and the antisymmetric tensor $h_{\alpha \beta } 
$ should be defined. In order to find the matrix $Z$ in this particular 
case, one should set up a Hermitian matrix $M$ written as

\begin{equation}
\label{eq20}
M = j^\alpha \cdot \left( {\tilde {\gamma }_\alpha \tilde {D}^{ - 1}} 
\right) + h^{\mu \nu } \cdot \left( {\tilde {S}_{\mu \nu } \tilde {D}^{ - 
1}} \right)
\end{equation}

\noindent
and extract a square root from it, i.e. find the matrix $Z$ satisfying the 
relationship

\begin{equation}
\label{eq21}
M = ZZ^ + .
\end{equation}

There are cases, when it is not difficult to extract a square root from $M$. 
Such cases include the situations, in which the matrix $M$ is positively 
defined, i.e. when all the eigenvalues $\left\{ {\mu _1 ,\;\mu _2 ,\;\mu _3 
,\;\mu _4 } \right\}$ are positive. Let us illustrate this. Let $\psi _A $ 
be an eigenvector of the matrix $M$ corresponding to the eigenvalue $\mu _A 
$ ($A = 1,2,3,4)$. By definition, the vectors $\psi _A $ satisfy the 
relationships

\begin{equation}
\label{eq22}
M\psi _A = \psi _A \mu _A .
\end{equation}

Each of the vectors $\psi _A $ has four components,

\begin{equation}
\label{eq23}
\psi _A = {\begin{array}{*{20}c}
 {\psi _{1A} } \hfill \\
 {\psi _{2A} } \hfill \\
 {\psi _{3A} } \hfill \\
 {\psi _{4A} } \hfill \\
\end{array} }.
\end{equation}

Since relationship (\ref{eq22}) defines the vectors $\psi _A $ to within the 
multiplication by a numerical factor, the vectors $\psi _A $ can be 
normalized by the condition

\begin{equation}
\label{eq24}
\left| {\psi _A } \right|^2 = \left| {\psi _{1A} } \right|^2 + \left| {\psi 
_{2A} } \right|^2 + \left| {\psi _{3A} } \right|^2 + \left| {\psi _{4A} } 
\right|^2 = 1.
\end{equation}

Let us introduce a matrix $U$, the columns in which consist of components of 
the normalized vectors $\psi _A $, i.e. the matrix

\begin{equation}
\label{eq25}
U = {\begin{array}{*{20}c}
 {\psi _{11} } \hfill & {\psi _{12} } \hfill & {\psi _{13} } \hfill & {\psi 
_{14} } \hfill \\
 {\psi _{21} } \hfill & {\psi _{22} } \hfill & {\psi _{23} } \hfill & {\psi 
_{24} } \hfill \\
 {\psi _{31} } \hfill & {\psi _{32} } \hfill & {\psi _{33} } \hfill & {\psi 
_{34} } \hfill \\
 {\psi _{41} } \hfill & {\psi _{42} } \hfill & {\psi _{43} } \hfill & {\psi 
_{44} } \hfill \\
\end{array} }.
\end{equation}

The matrix $U$ is unitary, i.e. $U^ + = U^{ - 1}$. We also introduce four 
projection operators $P_A $:

\begin{equation}
\label{eq26}
\begin{array}{l}
 P_1 = \frac{1}{2}\left( {E + \rho _3 } \right)\left( {E + \sigma _3 } 
\right),\;P_2 = \frac{1}{2}\left( {E + \rho _3 } \right)\left( {E - \sigma 
_3 } \right), \\ 
 P_3 = \frac{1}{2}\left( {E - \rho _3 } \right)\left( {E + \sigma _3 } 
\right),\;P_4 = \frac{1}{2}\left( {E - \rho _3 } \right)\left( {E - \sigma 
_3 } \right).\; \\ 
 \end{array}
\end{equation}

The set of projection operators (\ref{eq26}) is a full system in the sense that 
$\sum\limits_{A = 1,...,4} {P_A } = E$. Therefore, using the projection 
operators, matrix (\ref{eq25}) can be written as

\begin{equation}
\label{eq27}
U = \sum\limits_{A = 1,...,4} {\left( {UP_A } \right)} .
\end{equation}

If normalized matrix bispinors $u_A $ are introduced by means of the 
relationship

\begin{equation}
\label{eq28}
u_A \equiv UP_A ,
\end{equation}

Eq. (\ref{eq22}) will transform into

\begin{equation}
\label{eq29}
Mu_A = u_A \cdot \mu _A .
\end{equation}

Here, $\mu _A $ is a matrix\footnote{ Empty squares in the matrix contain 
zeros.} of the form

\begin{equation}
\label{eq30}
\mu _A = {\begin{array}{*{20}c}
 {\mu _1 } \hfill & \hfill & \hfill & \hfill \\
 \hfill & {\mu _2 } \hfill & \hfill & \hfill \\
 \hfill & \hfill & {\mu _3 } \hfill & \hfill \\
 \hfill & \hfill & \hfill & {\mu _4 } \hfill \\
\end{array} }.
\end{equation}

Summing of (\ref{eq29}) over $A$ gives:

\begin{equation}
\label{eq31}
MU = \sum\limits_{A = 1,...,4} {\left( {u_A \cdot \mu _A } \right)} = 
U\sum\limits_{A = 1,...,4} {\left( {\mu _A \cdot P_A } \right)} .
\end{equation}

Multiplication of (\ref{eq31}) in the right-hand member by $U^{ - 1}$ gives:

\begin{equation}
\label{eq32}
{\begin{array}{l}
M = U \cdot \sum\limits_{A = 1,...,4} {\left( {\mu _A P_A } \right)} \cdot 
U^+ =\\
= \left\{ {U \cdot \sum\limits_{A = 1,...,4} {\left( {\sqrt {\mu _A } 
\;P_A } \right)} } \right\} \cdot \left\{ {U \cdot \sum\limits_{A = 1,...,4} 
{\left( {\sqrt {\mu _A } \;P_A } \right)} } \right\}^+ = ZZ^+ .
\end{array}}
\end{equation}

As a result, we obtain relationship (\ref{eq21}), which needed to be proved. $Z$ is 
used here to denote the matrix

\begin{equation}
\label{eq33}
Z = U \cdot \left( {\sum\limits_{A = 1,...,4} {\left( {\sqrt {\mu _A } \;P_A 
} \right)} } \right).
\end{equation}

The explicit matrix form of the matrix $Z$ is

\begin{equation}
\label{eq34}
Z = U \cdot\left( \begin{array}{*{20}c}
 {\sqrt {\mu _1 } } \hfill & \hfill & \hfill & \hfill \\
 \hfill & {\sqrt {\mu _2 } } \hfill & \hfill & \hfill \\
 \hfill & \hfill & {\sqrt {\mu _3 } } \hfill & \hfill \\
 \hfill & \hfill & \hfill & {\sqrt {\mu _4 } }
\end{array}\right) .
\end{equation}

Note that eigenvalues of $M$ may also contain negative eigenvalues. In these 
cases, factorization of the matrix $M$ in the form of (\ref{eq21}) is inapplicable. 
Such cases are beyond the scope of this paper.

The matrix $Z$ in expansions of the type (\ref{eq21}) is defined ambiguously. If 
some matrix $Z_{ }$satisfies relationship (\ref{eq21}), the matrix ${Z}' = Z \cdot 
V$, where $V$ is an arbitrary unitary matrix, will satisfy the same 
relationship. Using this freedom, one can get the matrix $Z$ in (\ref{eq21}) to be 
Hermitian. It is easy to see that for this purpose it would suffice to 
assume that $V$ is equal to the matrix $U^{ - 1}$. 

\subsection{Theorem I}
\label{subsec:theorem}

Theorem I will be used to denote the relationship

\begin{equation}
\label{eq35}
j^\alpha _{;\alpha } = 0,
\end{equation}

\noindent
which can be easily checked by means of Eqs.(\ref{eq3}), (\ref{eq4}).

\subsection{Theorem II}
\label{subsec:mylabel2}

Theorem II states that if a bispinor matrix is governed by the Dirac 
equation (\ref{eq3}) and the quantities $j_\alpha $, $h_{\alpha \beta } $ are 
defined as (\ref{eq18}), (\ref{eq19}), the relationship

\begin{equation}
\label{eq36}
\left( {j_{\beta ,\alpha } - j_{\alpha ,\beta } } \right) = 4m \cdot 
h_{\alpha \beta } + E_{\alpha \beta \mu } ^\nu \frac{1}{4}\mbox{Sp}\left\{ 
{\left( {\nabla _\nu Z^ + } \right)\tilde {D}\tilde {\gamma }_5 \tilde 
{\gamma }^\mu Z - Z^ + \tilde {D}\tilde {\gamma }_5 \tilde {\gamma }^\mu 
\left( {\nabla _\nu Z} \right)} \right\}
\end{equation}

\noindent
holds.

The theorem will be proven in several steps. First, let us try to calculate 
the quantity $\left( {j_{\beta ,\alpha } - j_{\alpha ,\beta } } \right)$ 
based on relationships (\ref{eq12}), (\ref{eq13}). It follows from (\ref{eq18}) that:

\begin{equation}
\label{eq37}
\begin{array}{l}
 \left( {j_{\beta ,\alpha } - j_{\alpha ,\beta } } \right) = 
\frac{1}{4}\mbox{Sp}\left\{ {\left( {\nabla _\alpha Z^ + } \right)\tilde 
{D}\tilde {\gamma }_\beta Z + Z^ + \tilde {D}\tilde {\gamma }_\beta \left( 
{\nabla _\alpha Z} \right)} \right\} - \\ 
 - \frac{1}{4}\mbox{Sp}\left\{ {\left( {\nabla _\beta Z^ + } \right)\tilde 
{D}\tilde {\gamma }_\alpha Z + Z^ + \tilde {D}\tilde {\gamma }_\alpha \left( 
{\nabla _\beta Z} \right)} \right\}. \\ 
 \end{array}
\end{equation}

We use relationships (\ref{eq12}), (\ref{eq13}).

\begin{equation}
\label{eq38}
\begin{array}{l}
 \left( {j_{\beta ,\alpha } - j_{\alpha ,\beta } } \right) = \\ 
 = \frac{1}{4}\mbox{Sp}\left\{ \left( {\nabla _\nu Z^ + } \right)\tilde 
{D}\tilde {S}_\alpha ^\nu \tilde {\gamma }_\beta Z - m \cdot Z^ + \tilde 
{D}\tilde {\gamma }_\alpha \tilde {\gamma }_\beta Z - Z^+ \tilde {D}\tilde 
{\gamma }_\beta \tilde {S}_\alpha ^\nu \left( {\nabla _\nu Z} 
\right) + \right.  \\
\left. + m \cdot Z^ + \tilde {D}\tilde {\gamma }_\beta \tilde 
{\gamma }_\alpha Z 
\right\} 
 - \frac{1}{4}\mbox{Sp}\left\{ \left( {\nabla _\nu Z^+ } \right)\tilde 
{D}\tilde {S}_\beta^\nu \tilde {\gamma }_\alpha Z - \right. \\
\left. -m \cdot Z^ + \tilde{D}\tilde {\gamma }_\beta \tilde {\gamma }_\alpha Z - Z^ + \tilde {D}\tilde 
{\gamma }_\alpha \tilde {S}_\beta ^\nu \left( {\nabla _\nu Z} \right. + m 
\cdot Z^ + \tilde {D}\tilde {\gamma }_\alpha \tilde {\gamma }_\beta Z 
\right\}. \\ 
 \end{array}
\end{equation}

Terms without derivatives and with derivatives are grouped separately in the 
right-hand member of (\ref{eq38}).

\begin{equation}
\label{eq39}
\begin{array}{l}
 \left( {j_{\beta ,\alpha } - j_{\alpha ,\beta } } \right) = \frac{1}{2}m 
\cdot \mbox{Sp}\left\{ { - Z^ + \tilde {D}\tilde {\gamma }_\alpha \tilde 
{\gamma }_\beta Z + Z^ + \tilde {D}\tilde {\gamma }_\beta \tilde {\gamma 
}_\alpha Z} \right\} \\ 
 + \frac{1}{4}\mbox{Sp}\left\{ {\left( {\nabla _\nu Z^ + } \right)\tilde 
{D}\tilde {S}_\alpha ^\nu \tilde {\gamma }_\beta Z - Z^ + \tilde {D}\tilde 
{\gamma }_\beta \tilde {S}_\alpha ^\nu \left( {\nabla _\nu Z} \right) - 
\left( {\nabla _\nu Z^ + } \right)\tilde {D}\tilde {S}_\beta ^\nu \tilde 
{\gamma }_\alpha Z }\right.\\
\left. + Z^+ \tilde {D}\tilde {\gamma }_\alpha \tilde {S}_\beta 
^\nu \left( {\nabla _\nu Z} \right) \right\}. \\ 
 \end{array}
\end{equation}

DM products, such as $\tilde {S}_\alpha ^\nu \tilde {\gamma }_\beta $, 
$\tilde {\gamma }_\beta \tilde {S}_\alpha ^\nu $, are replaced in (\ref{eq39}) with 
the following relationships:

\[
\left. {\begin{array}{l}
 \tilde {S}_\alpha ^\nu \tilde {\gamma }_\beta = - \eta _{\alpha \beta } 
\tilde {\gamma }^\nu + \delta _\beta ^\nu \tilde {\gamma }_\alpha + E_\alpha 
{}^\nu _{\beta \mu } \tilde {\gamma }_5 \tilde {\gamma }^\mu \\ 
 \tilde {\gamma }_\beta \tilde {S}_\alpha ^\nu = \eta _{\alpha \beta } 
\tilde {\gamma }^\nu - \delta _\beta ^\nu \tilde {\gamma }_\alpha + E_\alpha 
{}^\nu _{\beta \mu } \tilde {\gamma }_5 \tilde {\gamma }^\mu \\ 
 \end{array}} \right\} .
\]

We have:

\begin{equation}
\label{eq40}
\begin{array}{l}
 \left( {j_{\beta ,\alpha } - j_{\alpha ,\beta } } \right) = - m \cdot 
\mbox{Sp}\left\{ {Z^ + \tilde {D}\tilde {S}_{\alpha \beta } Z} \right\} \\ 
 + \frac{1}{4}\mbox{Sp}\left\{ {\left( {\nabla _\nu Z^ + } \right)\tilde 
{D}\left( { - \eta _{\alpha \beta } \tilde {\gamma }^\nu + \delta _\beta 
{}^\nu \tilde {\gamma }_\alpha + E_{\alpha} {}^{\nu}_{\beta \mu } \tilde {\gamma 
}_5 \tilde {\gamma }^\mu } \right)Z} \right\} \\ 
 + \frac{1}{4}\mbox{Sp}\left\{ { - Z^ + \tilde {D}\left( {\eta _{\alpha 
\beta } \tilde {\gamma }^\nu - \delta _\beta ^\nu \tilde {\gamma }_\alpha + 
E_\alpha {}^\nu _{\beta \mu } \tilde {\gamma }_5 \tilde {\gamma }^\mu } 
\right)\left( {\nabla _\nu Z} \right)} \right\} \\ 
 + \frac{1}{4}\mbox{Sp}\left\{ { - \left( {\nabla _\nu Z^ + } \right)\tilde 
{D}\left( { - \eta _{\alpha \beta } \tilde {\gamma }^\nu + \delta _\alpha 
{}^\nu \tilde {\gamma }_\beta - E_\alpha {}^\nu _{\beta \mu } \tilde {\gamma }_5 
\tilde {\gamma }^\mu } \right)Z} \right\} \\ 
 + \frac{1}{4}\mbox{Sp}\left\{ { + Z^ + \tilde {D}\left( {\eta _{\alpha 
\beta } \tilde {\gamma }^\nu - \delta _\alpha ^\nu \tilde {\gamma }_\beta - 
E_\alpha {}^\nu _{\beta \mu } \tilde {\gamma }_5 \tilde {\gamma }^\mu } 
\right)\left( {\nabla _\nu Z} \right)} \right\}. \\ 
 \end{array}
\end{equation}

The terms with the metric tensor in (\ref{eq40}) are canceled. The rest give:

\[
\begin{array}{l}
 \left( {j_{\beta ,\alpha } - j_{\alpha ,\beta } } \right) = + 8m \cdot 
h_{\alpha \beta } \\ 
 + \frac{1}{4}\mbox{Sp}\left\{ {\left( {\nabla _\beta Z^ + } \right)\tilde 
{D}\tilde {\gamma }_\alpha Z} \right\} + E_\alpha {}^\nu _{\beta \mu } 
\frac{1}{4}\mbox{Sp}\left\{ {\left( {\nabla _\nu Z^ + } \right)\tilde 
{D}\tilde {\gamma }_5 \tilde {\gamma }^\mu Z} \right\} \\ 
 + \frac{1}{4}\mbox{Sp}\left\{ {Z^ + \tilde {D}\tilde {\gamma }_\alpha 
\left( {\nabla _\beta Z} \right)} \right\} - E_\alpha {}^\nu _{\beta \mu } 
\frac{1}{4}\mbox{Sp}\left\{ {Z^ + \tilde {D}\tilde {\gamma }_5 \tilde 
{\gamma }^\mu \left( {\nabla _\nu Z} \right)} \right\} \\ 
 + \frac{1}{4}\mbox{Sp}\left\{ { - \left( {\nabla _\alpha Z^ + } 
\right)\tilde {D}\tilde {\gamma }_\beta Z} \right\} + E_\alpha {}^\nu _{\beta 
\mu } \frac{1}{4}\mbox{Sp}\left\{ {\left( {\nabla _\nu Z^ + } \right)\tilde 
{D}\tilde {\gamma }_5 \tilde {\gamma }^\mu Z} \right\} \\ 
 + \frac{1}{4}\mbox{Sp}\left\{ { - Z^ + \tilde {D}\tilde {\gamma }_\beta 
\left( {\nabla _\alpha Z} \right)} \right\} - E_\alpha {}^\nu _{\beta \mu } 
\frac{1}{4}\mbox{Sp}\left\{ { + Z^ + \tilde {D}\tilde {\gamma }_5 \tilde 
{\gamma }^\mu \left( {\nabla _\nu Z} \right)} \right\}. \\ 
 \end{array}
\]

The terms with $\tilde {D}\tilde {\gamma }_\alpha $, $\tilde {D}\tilde 
{\gamma }_\alpha $ are reduced to $ - \left( {j_{\beta ,\alpha } - j_{\alpha 
,\beta } } \right)$. The rest are combined into the expression $E_\alpha 
{}^\nu _{\beta \mu } \frac{1}{2}\mbox{Sp}\left\{ {\left( {\nabla _\nu Z^ + } 
\right)\tilde {D}\tilde {\gamma }_5 \tilde {\gamma }^\mu Z - Z^ + \tilde 
{D}\tilde {\gamma }_5 \tilde {\gamma }^\mu \left( {\nabla _\nu Z} \right)} 
\right\}$. As a result, we have

\begin{equation}
\label{eq41}
\begin{array}{l}
\left( {j_{\beta ,\alpha } - j_{\alpha ,\beta } } \right) = 8m \cdot 
h_{\alpha \beta } - \left( {j_{\beta ,\alpha } - j_{\alpha ,\beta } } 
\right) \\
+ E_\alpha {}^\nu _{\beta \mu } \frac{1}{2}\mbox{Sp}\left\{ {\left( 
{\nabla _\nu Z^ + } \right)\tilde {D}\tilde {\gamma }_5 \tilde {\gamma }^\mu 
Z - Z^ + \tilde {D}\tilde {\gamma }_5 \tilde {\gamma }^\mu \left( {\nabla 
_\nu Z} \right)} \right\}.\\
\end{array}
\end{equation}

Identical transformations of relationship (\ref{eq41}) give the same expression as 
(\ref{eq36}). Thus, Theorem II is proven. 

\subsection{Theorem III}
\label{subsec:mylabel3}

If the bispinor matrix $Z$ satisfies the Dirac equation (\ref{eq3}), the quantity 
$h_{\alpha \beta } $ defined by Eq. (\ref{eq19}) will satisfy the relationship

\begin{equation}
\label{eq42}
E^{\lambda \alpha \beta \varepsilon }h_{\alpha \beta ;\varepsilon } = 
\frac{1}{4} \cdot \mbox{Sp}\left\{ {\left( {\nabla ^\lambda Z^ + } 
\right)\tilde {D}\tilde {\gamma }_5 Z - Z^ + \tilde {D}\tilde {\gamma }_5 
\left( {\nabla ^\lambda Z} \right)} \right\}.
\end{equation}

Equality (\ref{eq42}) will be below referred to as Theorem III.

To prove it, let us multiply (\ref{eq36}) by $E^{\lambda \alpha \beta \varepsilon }$ 
and differentiate it in the index $\varepsilon $.

\begin{equation}
\label{eq43}
\begin{array}{l}
 E^{\lambda \alpha \beta \varepsilon }\left( {j_{\beta ;\alpha } - j_{\alpha 
;\beta } } \right)_{;\varepsilon } = 4m \cdot E^{\lambda \alpha \beta 
\varepsilon }h_{\alpha \beta ;\varepsilon } \\ 
 + E^{\lambda \alpha \beta \varepsilon }E_{\alpha \beta \mu } ^\nu 
\frac{1}{4}\mbox{Sp}\left\{ {\left( {\nabla _\nu Z^ + } \right)\tilde 
{D}\tilde {\gamma }_5 \tilde {\gamma }^\mu Z - Z^ + \tilde {D}\tilde {\gamma 
}_5 \tilde {\gamma }^\mu \left( {\nabla _\nu Z} \right)} 
\right\}_{;\varepsilon } \\ 
 \end{array}.
\end{equation}

The term in the left-hand member of (\ref{eq43}) identically vanishes. In order to 
transform the right-hand member, we use the formula

\[
E^{\lambda \alpha \beta \varepsilon }E_{\alpha \beta \mu } ^\nu = - 2\delta 
_\mu ^\lambda \eta ^{\varepsilon \nu } + 2\delta _\mu ^\varepsilon \eta 
^{\lambda \nu }.
\]

As a result:

\begin{equation}
\label{eq44}
\begin{array}{l}
 - 4m \cdot E^{\lambda \alpha \beta \varepsilon }h_{\alpha \beta 
;\varepsilon } = - \frac{1}{2}\mbox{Sp}\left\{ {\left( {\nabla _\nu Z^ + } 
\right)\tilde {D}\tilde {\gamma }_5 \tilde {\gamma }^\lambda Z - Z^ + \tilde 
{D}\tilde {\gamma }_5 \tilde {\gamma }^\lambda \left( {\nabla _\nu Z} 
\right)} \right\}^{;\nu } \\ 
 + \frac{1}{2}\mbox{Sp}\left\{ {\left( {\nabla ^\lambda Z^ + } \right)\tilde 
{D}\tilde {\gamma }_5 \tilde {\gamma }^\mu Z - Z^ + \tilde {D}\tilde {\gamma 
}_5 \tilde {\gamma }^\mu \left( {\nabla ^\lambda Z} \right)} \right\}_{;\mu 
} . \\ 
 \end{array}.
\end{equation}

Derivatives in (\ref{eq44}) are developed.

\begin{equation}
\label{eq45}
\begin{array}{l}
 - 4m \cdot E^{\lambda \alpha \beta \varepsilon }h_{\alpha \beta 
;\varepsilon } = - \frac{1}{2}\mbox{Sp}\left\{ {\left( {\nabla ^\nu \nabla 
_\nu Z^ + } \right)\tilde {D}\tilde {\gamma }_5 \tilde {\gamma }^\lambda Z - 
Z^ + \tilde {D}\tilde {\gamma }_5 \tilde {\gamma }^\lambda \left( {\nabla 
^\nu \nabla _\nu Z} \right)} \right\} \\ 
 + \frac{1}{2}\mbox{Sp}\left\{ {\left( {\nabla _\mu \nabla ^\lambda Z^ + } 
\right)\tilde {D}\tilde {\gamma }_5 \tilde {\gamma }^\mu Z - Z^ + \tilde 
{D}\tilde {\gamma }_5 \tilde {\gamma }^\mu \left( {\nabla _\mu \nabla 
^\lambda Z} \right)} \right\} \\ 
 + \frac{1}{2}\mbox{Sp}\left\{ {\left( {\nabla ^\lambda Z^ + } \right)\tilde 
{D}\tilde {\gamma }_5 \tilde {\gamma }^\mu \left( {\nabla _\mu Z} \right) - 
\left( {\nabla _\mu Z^ + } \right)\tilde {D}\tilde {\gamma }_5 \tilde 
{\gamma }^\mu \left( {\nabla ^\lambda Z} \right)} \right\} = \\ 
 = - \frac{1}{2}\mbox{Sp}\left\{ {\left( {\nabla ^\nu \nabla _\nu Z^ + } 
\right)\tilde {D}\tilde {\gamma }_5 \tilde {\gamma }^\lambda Z - Z^ + \tilde 
{D}\tilde {\gamma }_5 \tilde {\gamma }^\lambda \left( {\nabla ^\nu \nabla 
_\nu Z} \right)} \right\} \\ 
 + \frac{1}{2}\mbox{Sp}\left\{ {\left( {\nabla _\mu \nabla ^\lambda Z^ + } 
\right)\tilde {D}\tilde {\gamma }_5 \tilde {\gamma }^\mu Z - Z^ + \tilde 
{D}\tilde {\gamma }_5 \tilde {\gamma }^\mu \left( {\nabla _\mu \nabla 
^\lambda Z} \right)} \right\} \\ 
 + \frac{1}{2}m \cdot \mbox{Sp}\left\{ {\left( {\nabla ^\lambda Z^ + } 
\right)\tilde {D}\tilde {\gamma }_5 Z - Z^ + \tilde {D}\tilde {\gamma }_5 
\left( {\nabla ^\lambda Z} \right)} \right\}. \\ 
 \end{array}.
\end{equation}

Relationship (\ref{eq45}) can be simplified if we use relationships (\ref{eq9}) and (\ref{eq17}).

\begin{equation}
\label{eq46}
\begin{array}{l}
 4m \cdot E^{\lambda \alpha \beta \varepsilon }h_{\alpha \beta ;\varepsilon 
} = - \frac{1}{2}\mbox{Sp}\left\{ {\frac{1}{2}P_{\mu \nu } Z^ + \tilde 
{D}\tilde {\gamma }_5 \tilde {S}^{\mu \nu }\gamma ^\lambda Z - \frac{1}{2}Z^ 
+ \tilde {D}\tilde {\gamma }_5 \tilde {\gamma }^\lambda \tilde {S}^{\mu \nu 
}ZP_{\mu \nu } } \right\} \\ 
 + \frac{1}{2}\mbox{Sp}\left\{ { - \left( {\nabla ^\lambda \nabla _\mu Z^ + 
} \right)\tilde {D}\tilde {\gamma }^\mu \tilde {\gamma }_5 Z - Z^ + \tilde 
{D}\tilde {\gamma }_5 \tilde {\gamma }^\mu \left( {\nabla ^\lambda \nabla 
_\mu Z} \right) - 2P^\lambda _\mu \cdot \left( {Z^ + \tilde {D}\tilde 
{\gamma }_5 \tilde {\gamma }^\mu Z} \right)} \right\} \\ 
 + \frac{1}{2}m \cdot \mbox{Sp}\left\{ {\left( {\nabla ^\lambda Z^ + } 
\right)\tilde {D}\tilde {\gamma }_5 Z - Z^ + \tilde {D}\tilde {\gamma }_5 
\left( {\nabla ^\lambda Z} \right)} \right\}. \\ 
 \end{array}.
\end{equation}

The terms with second derivatives in (\ref{eq46}) are transformed.

\begin{equation}
\label{eq47}
E^{\lambda \alpha \beta \varepsilon }h_{\alpha \beta ;\varepsilon } = 
\frac{1}{4} \cdot \mbox{Sp}\left\{ {\left( {\nabla ^\lambda Z^ + } 
\right)\tilde {D}\tilde {\gamma }_5 Z - Z^ + \tilde {D}\tilde {\gamma }_5 
\left( {\nabla ^\lambda Z} \right)} \right\}.
\end{equation}

Thus, we obtain an equality, which coincides with the statement of Theorem 
III, i.e. with equality (\ref{eq42}).

\subsection{Theorem IV}
\label{subsec:mylabel4}

Theorem IV proves that the relationship

\begin{equation}
\label{eq48}
h_\alpha {}^\nu _{;\nu } = - \frac{1}{8}\mbox{Sp}\left( {\left( {\nabla 
_\alpha Z^ + } \right)\tilde {D}Z - Z^ + \tilde {D}\left( {\nabla _\alpha Z} 
\right)} \right) - m \cdot j_\alpha 
\end{equation}

\noindent
holds.

The proof will consist in the direct validation of (\ref{eq48}).

\begin{equation}
\label{eq49}
\begin{array}{l}
 h_\alpha {}^\nu _{;\nu } = - \frac{1}{8}\mbox{Sp}\left\{ {Z^ + \tilde 
{D}\tilde {S}_\alpha ^\nu Z} \right\}_{;\nu } = \\ 
 = - \frac{1}{8}\mbox{Sp}\left\{ {\left( {\nabla _\nu Z^ + } \right)\tilde 
{D}\tilde {S}_\alpha ^\nu Z} \right\} - \frac{1}{8}\mbox{Sp}\left\{ {Z^ + 
\tilde {D}\tilde {S}_\alpha ^\nu \left( {\nabla _\nu Z} \right)} \right\}. 
\\ 
 \end{array}
\end{equation}

The matrix $\tilde {S}_\alpha ^\nu $ in the first case is replaced with

\begin{equation}
\label{eq50}
\tilde {S}_\alpha ^\nu = \delta _\alpha ^\nu - \tilde {\gamma }^\nu \tilde 
{\gamma }_\alpha ,
\end{equation}

\noindent
and in the second case, with

\begin{equation}
\label{eq51}
\tilde {S}_\alpha ^\nu = - \delta _\alpha ^\nu + \tilde {\gamma }_\alpha 
\tilde {\gamma }^\nu .
\end{equation}

Hence:

\begin{equation}
\label{eq52}
\begin{array}{l}
 h_\alpha {}^\nu _{;\nu } = - \frac{1}{8}\mbox{Sp}\left\{ {\left( {\nabla _\nu 
Z^ + } \right)\tilde {D}\left( {\delta _\alpha ^\nu - \tilde {\gamma }^\nu 
\tilde {\gamma }_\alpha } \right)Z} \right\} \\
- \frac{1}{8}\mbox{Sp}\left\{ 
{Z^ + \tilde {D}\left( { - \delta _\alpha ^\nu + \tilde {\gamma }_\alpha 
\tilde {\gamma }^\nu } \right)\left( {\nabla _\nu Z} \right)} \right\} = \\ 
 = - \frac{1}{8}\mbox{Sp}\left\{ {\left( {\nabla _\nu Z^ + } \right)\tilde 
{D}\left( {\delta _\alpha ^\nu } \right)Z} \right\} - 
\frac{1}{8}\mbox{Sp}\left\{ {\left( {\nabla _\nu Z^ + } \right)\tilde 
{D}\left( { - \tilde {\gamma }^\nu \tilde {\gamma }_\alpha } \right)Z} 
\right\} \\ 
 - \frac{1}{8}\mbox{Sp}\left\{ {Z^ + \tilde {D}\left( { - \delta _\alpha 
^\nu } \right)\left( {\nabla _\nu Z} \right)} \right\} - 
\frac{1}{8}\mbox{Sp}\left\{ {Z^ + \tilde {D}\left( {\tilde {\gamma }_\alpha 
\tilde {\gamma }^\nu } \right)\left( {\nabla _\nu Z} \right)} \right\} = \\ 
 = - \frac{1}{8}\mbox{Sp}\left\{ {\left( {\nabla _\alpha Z^ + } 
\right)\tilde {D}Z - Z^ + \tilde {D}\left( {\nabla _\alpha Z} \right)} 
\right\} \\ 
 + \frac{1}{8}\mbox{Sp}\left\{ {\left( {\nabla _\nu Z^ + } \right)\tilde 
{D}\tilde {\gamma }^\nu \tilde {\gamma }_\alpha Z} \right\} - 
\frac{1}{8}\mbox{Sp}\left\{ {Z^ + \tilde {D}\tilde {\gamma }_\alpha \tilde 
{\gamma }^\nu \left( {\nabla _\nu Z} \right)} \right\} \\ 
 \end{array}
\end{equation}

Using Eqs. (\ref{eq3}), (\ref{eq4}), we obtain:

\begin{equation}
\label{eq53}
h_\alpha {}^\nu _{;\nu } = \frac{1}{8}\mbox{Sp}\left\{ {\left( {\nabla _\alpha 
Z^ + } \right)\tilde {D}Z - Z^ + \tilde {D}\left( {\nabla _\alpha Z} 
\right)} \right\} - 2m \cdot \frac{1}{8}\mbox{Sp}\left\{ {Z^ + \tilde 
{D}\tilde {\gamma }_\alpha Z} \right\}.
\end{equation}

Thus, it is proven that relationship (\ref{eq48}) indeed follows from the Dirac 
equation, i.e. Theorem IV is proven.

\section{CGD Equations in Plane Space}
\label{sec:mylabel1}

Conformal geometrodynamics equations in the general case are equations

\begin{equation}
\label{eq54}
R_{\alpha \beta } - \frac{1}{2}g_{\alpha \beta } R = - 2A_\alpha A_\beta - 
g_{\alpha \beta } A^2 - 2g_{\alpha \beta } A^\nu _{;\nu } + A_{\alpha ;\beta 
} + A_{\beta ;\alpha } + \lambda g_{\alpha \beta } .
\end{equation}

The right-hand member in Eqs. (\ref{eq54}), i.e. the tensor

\begin{equation}
\label{eq55}
T_{\alpha \beta } = - 2A_\alpha A_\beta - g_{\alpha \beta } A^2 - 2g_{\alpha 
\beta } A^\nu _{;\nu } + A_{\alpha ;\beta } + A_{\beta ;\alpha } + \lambda 
g_{\alpha \beta } ,
\end{equation}

\noindent
is considered to be the energy-momentum tensor of geometrodynamic continuum. 
A corollary of Eq. (\ref{eq54}) and Bianchi identities is that

\begin{equation}
\label{eq56}
T_\alpha {}^\beta _{;\beta } = 0.
\end{equation}

It follows from Eqs. (\ref{eq56}) that the tensor $F_{\alpha \beta } $ defined as

\begin{equation}
\label{eq57}
F_{\alpha \beta } = A_{\beta ,\alpha } - A_{\alpha ,\beta } ,
\end{equation}

\noindent
should satisfy the equation

\begin{equation}
\label{eq58}
F_{\alpha \cdot } {}^\beta _{;\beta } = \lambda _{;\alpha } - 2\lambda 
A_\alpha .
\end{equation}

Let us express all these relationships for plane space in Cartesian 
coordinates chosen as world coordinates for the case, when the metric tensor 
is given by

\begin{equation}
\label{eq59}
g_{\alpha \beta } \equiv \eta _{\alpha \beta } = \mbox{diag}\left[ { - 
1,1,1,1} \right],
\end{equation}

We write the full set of dynamic equations and the gauge condition, which 
will be solved below for the flat Riemann space:

\begin{equation}
\label{eq60}
\left. {\begin{array}{l}
 A_{\beta ,\alpha } - A_{\alpha ,\beta } = F_{\alpha \beta } \\ 
 F_\alpha {}^\beta _{;\beta } = \lambda _{,\alpha } - 2\lambda A_\alpha \\ 
 \lambda = \mbox{Const},\quad A^\nu _{;\nu } = 0 \\ 
 g_{\alpha \beta } \equiv \eta _{\alpha \beta } \\ 
 \end{array}} \right\} .
\end{equation}

The gauge condition here is

\begin{equation}
\label{eq61}
\lambda = \mbox{Const}.
\end{equation}

A special feature of condition (\ref{eq61}) is that when it is fulfilled, the 
relationship

\begin{equation}
\label{eq62}
A^\nu _{;\nu } = 0,
\end{equation}

\noindent
which is also part of (\ref{eq60}), is fulfilled as well.

In principle, one can exclude $F_{\alpha \beta } $'s$_{ }$from (\ref{eq60}) and 
reduce all the equations to the second-order equation for $A_\alpha $ 
supplemented with the gauge condition. We will use a different approach, in 
which all the eleven functions of $A_\alpha $, $F_{\alpha \beta } $, 
$\lambda $ versus four variables are treated as independent, and equations 
(\ref{eq60}) are treated as a set of eleven first-order partial differential 
equations for these functions. Such an approach is similar to the 
representation of the Klein-Gordon equations as a set of first-order 
equations, see Section 4.4 in \cite{Bogo}.

A direct check can prove that for each solution of the CGD equations one can 
introduce a dimensionless vector

\begin{equation}
\label{eq63}
J^\alpha \equiv \frac{\kappa }{\lambda ^{3 / 2}}\left( {\nabla ^\alpha 
\lambda } \right) = \frac{\kappa }{\lambda ^{3 / 2}}g^{\alpha \beta }\left( 
{\lambda _{;\beta } - 2\lambda A_\beta } \right)
\end{equation}

\noindent
with the Weyl weight of -1 and a dimensionless anti-symmetric tensor

\begin{equation}
\label{eq64}
H^{\alpha \beta } \equiv \frac{\theta }{\lambda }F^{\alpha \beta } = 
\frac{\theta }{\lambda }g^{\alpha \mu }g^{\beta \nu }\left( {A_{\nu ,\mu } - 
A_{\mu ,\nu } } \right)
\end{equation}

\noindent
with the Weyl weight of -2. The quantities $\kappa $ in (\ref{eq63}) and $\theta $ 
in (\ref{eq64}) are some constant coefficients.

As applied to the vector $J_\alpha $, gauge condition (\ref{eq60}) is written as

\begin{equation}
\label{eq65}
J^\alpha _{;\alpha } = 0.
\end{equation}

Since $\lambda = \mbox{Const}_{ }$in such gauging, the vector $J_\alpha $ 
and the tensor $H_{\alpha \beta } $ are related to the initial quantities 
$A_\alpha $, $F_{\alpha \beta } $, $\lambda $ as

\begin{equation}
\label{eq66}
A_\alpha = - \frac{\lambda ^{1 / 2}}{2\kappa }J_\alpha ;\quad F_{\alpha 
\beta } = \frac{\lambda }{\theta }H_{\alpha \beta } ,
\end{equation}

\noindent
and to each other, as

\begin{equation}
\label{eq67}
\left( {J_{\beta ;\alpha } - J_{\alpha ;\beta } } \right) = - \frac{2\kappa 
}{\theta }\lambda ^{1 / 2}H_{\alpha \beta } ,
\end{equation}

\begin{equation}
\label{eq68}
H_{\alpha \cdot } {}^\beta _{;\beta } = \frac{\theta }{\kappa }\lambda ^{1 / 
2} \cdot J_\alpha .
\end{equation}

It follows from equality (\ref{eq67}) that the tensor$H_{\alpha \beta } $ satisfies 
four identities

\begin{equation}
\label{eq69}
E^{\alpha \beta \mu \nu }H_{\beta \mu ,\nu } \equiv 0.
\end{equation}

Consider the Cauchy problem for the vector $J_\alpha $ and tensor $H_{\alpha 
\beta } $. For this purpose, relationships (\ref{eq65})-(\ref{eq69}) are written in 
Cartesian coordinates with differentiation of space and time indices.

Dynamic equations:

\begin{equation}
\label{eq70}
\left[ {J_{k,0} - J_{0,k} } \right] = - \frac{2\kappa }{\theta }\lambda ^{1 
/ 2}H_{0k} ,
\end{equation}

\begin{equation}
\label{eq71}
H_{0k,0} + H_{kp,p} = \frac{\theta }{\kappa }\lambda ^{1 / 2}\; \cdot J_k ,
\end{equation}

\begin{equation}
\label{eq72}
J_{0,0} = J_{p,p} ,
\end{equation}

\begin{equation}
\label{eq73}
H_{pq,0} = - H_{0p,q} + H_{0q,p} .
\end{equation}

Coupling equations:

\begin{equation}
\label{eq74}
\left[ {J_{n,m} - J_{m,n} } \right] = - \frac{2\kappa }{\theta }\lambda ^{1 
/ 2}H_{mn} ,
\end{equation}

\begin{equation}
\label{eq75}
H_{0p,p} = \frac{\theta }{\kappa }\lambda ^{1 / 2}\; \cdot J_0 ,
\end{equation}

\begin{equation}
\label{eq76}
H_{pq,k} = - H_{kp,q} + H_{kq,p} .
\end{equation}

Due to their structure, relationships (\ref{eq70})-(\ref{eq73}) can be considered a set of 
first-order differential equations with respect to the desired functions 
$J_0 $, $J_k $, $H_{0k} $, $H_{mn} $. These functions are basically Cauchy 
data. Indeed, time derivatives of these functions are defined by dynamic 
equations (\ref{eq70}), (\ref{eq71}), (\ref{eq72}), (\ref{eq73}). As for relationships (\ref{eq74}), (\ref{eq75}), (\ref{eq76}), 
these are connections to the Cauchy data. Relationship (\ref{eq76}), which is 
fulfilled automatically if relationship (\ref{eq74}) is fulfilled, can be excluded 
from the list of connections.

Such a statement of the Cauchy problem would be correct if connections (\ref{eq74}), 
(\ref{eq75}) were "drawn'' into volume by virtue of dynamic equations (\ref{eq70}), (\ref{eq71}), 
(\ref{eq72}), (\ref{eq73}). It is easy to show (we will not do this here) that this 
situation applies.

\section{Conditions of Coincidence of CGD Equations and Dirac Equation Corollaries}
\label{sec:conditions}

\subsection{Comparison of Equations }

\ref{tab1} shows the CGD equations and respective 
equations, which are direct corollaries of the Dirac equation.

\noindent
\begin{table}[htbp]
\begin{tabular}
{|p{271pt}|p{92pt}|}
\hline
Diract equation corollaries& 
CGD equations \\
\hline
$j^\alpha _{;\alpha } = 0$ \par Relationship (\ref{eq35})& 
$J^\alpha _{;\alpha } = 0$ \par Relationship (\ref{eq65}) \\
\hline
$\begin{array}{l}
 \left( {j_{\beta ,\alpha } - j_{\alpha ,\beta } } \right) =\\
 4m \cdot h_{\alpha \beta } + \\ 
 + E_{\alpha \beta \mu } ^\nu \frac{1}{4}\mbox{Sp}\left\{ {\left( 
{\nabla _\nu Z^ + } \right)\tilde {D}\tilde {\gamma }_5 \tilde 
{\gamma }^\mu Z - Z^ + \tilde {D}\tilde {\gamma }_5 \tilde 
{\gamma }^\mu \left( {\nabla _\nu Z} \right)} \right\} \\ 
 \end{array}$ \par Relationships (\ref{eq36})&
$\begin{array}{l}
\left( {J_{\beta ;\alpha } - J_{\alpha ;\beta } } \right)\\ 
= - \frac{2\kappa }{\theta }\lambda ^{1 / 2}H_{\alpha \beta } \\
\end{array}$ 
\par Relationships (\ref{eq67}) \\
\hline
$\begin{array}{l}
 h_\alpha {}^\nu _{;\nu } = - m \cdot j_\alpha \\ 
 - \frac{1}{8}\mbox{Sp}\left( {\left( {\nabla _\alpha Z^ + } \right)\tilde {D}Z - Z^ + \tilde {D}\left( {\nabla _\alpha Z} \right)} \right) \\ 
 \end{array}$ \par Equations (\ref{eq48})& 
$H_{\alpha \cdot } {}^\beta _{;\beta } = \frac{\theta }{\kappa }\lambda^{1 / 2} \cdot J_\alpha $ \par Equations (\ref{eq68}) \\
\hline
$E^{\lambda \alpha \beta \varepsilon }h_{\alpha \beta ;\varepsilon } 
= \frac{1}{4} \cdot \mbox{Sp}\left\{ {\left( {\nabla ^\lambda Z^+ } 
\right)\tilde {D}\tilde {\gamma }_5 Z - Z^ + \tilde {D}\tilde {\gamma }_5 \left( {\nabla ^\lambda Z} \right)} \right\}$ \par Equations (\ref{eq42})& 
$E^{\alpha \beta \mu \nu }H_{\beta \mu ,\nu } \equiv 0$ \par Equations (\ref{eq69}) \\
\hline
\end{tabular}
\caption{-- Comparison of CGD equations with Dirac equation corollaries}
\label{tab1}
\end{table}

Let us question ourselves, what conditions should be satisfied for the 
corresponding (i.e. given in the same lines of \ref{tab1}) equations to coincide. It is clear that the following relationships 
should be fulfilled in the first place:

\begin{equation}
\label{eq77}
H_{\alpha \beta } = h_{\alpha \beta } ,\quad J_\alpha = j_\alpha ,
\end{equation}

\noindent
and in the second place:

\begin{equation}
\label{eq78}
m = - \left( {\kappa \mathord{\left/ {\vphantom {\kappa {2\theta }}} \right. 
\kern-\nulldelimiterspace} {2\theta }} \right)\lambda ^{1 / 2},\quad m = - 
\left( {\theta \mathord{\left/ {\vphantom {\theta \kappa }} \right. 
\kern-\nulldelimiterspace} \kappa } \right)\lambda ^{1 / 2}.
\end{equation}

Conditions (\ref{eq78}) are satisfied if the following relationship between the 
coefficients $\kappa $ and $\theta $ is fulfilled:

\begin{equation}
\label{eq79}
\kappa = - \theta \sqrt 2 .
\end{equation}

The mass and the quantity $\lambda $ should be related as

\begin{equation}
\label{eq80}
m = \sqrt {\lambda \mathord{\left/ {\vphantom {\lambda 2}} \right. 
\kern-\nulldelimiterspace} 2} .
\end{equation}

The third condition is that the following three quantities should 
vanish:

\begin{equation}
\label{eq81}
E_{\alpha \beta \mu } ^\nu \mbox{Sp}\left\{ {\left( {\nabla _\nu Z^ + } 
\right)\tilde {D}\tilde {\gamma }_5 \tilde {\gamma }^\mu Z - Z^ + \tilde 
{D}\tilde {\gamma }_5 \tilde {\gamma }^\mu \left( {\nabla _\nu Z} \right)} 
\right\} = 0,
\end{equation}

\begin{equation}
\label{eq82}
\mbox{Sp}\left( {\left( {\nabla _\alpha Z^ + } \right)\tilde {D}Z - Z^ + 
\tilde {D}\left( {\nabla _\alpha Z} \right)} \right) = 0,
\end{equation}

\begin{equation}
\label{eq83}
\mbox{Sp}\left\{ {\left( {\nabla ^\lambda Z^ + } \right)\tilde {D}\tilde 
{\gamma }_5 Z - Z^ + \tilde {D}\tilde {\gamma }_5 \left( {\nabla ^\lambda Z} 
\right)} \right\} = 0.
\end{equation}

In order to ascertain the meaning of relationships (\ref{eq81}) - (\ref{eq83}), we 
introduce a matrix

\begin{equation}
\label{eq84}
n_\alpha \equiv Z^{ - 1}Z_{,\alpha } .
\end{equation}

These relationships will change to:

\begin{equation}
\label{eq85}
\;\mbox{Sp}\left\{ {\Gamma _\alpha D\gamma _5 \gamma ^\mu } \right\} = 
\mbox{Sp}\left\{ {\frac{1}{2}\left( {n_\alpha - n_\alpha ^ + } 
\right)D\gamma _5 \gamma ^\mu } \right\},
\end{equation}

\begin{equation}
\label{eq86}
\;\mbox{Sp}\left\{ {\Gamma _\alpha D} \right\} = \mbox{Sp}\left\{ 
{\frac{1}{2}\left( {n_\alpha - n_\alpha ^ + } \right)D} \right\},
\end{equation}

\begin{equation}
\label{eq87}
\;\mbox{Sp}\left\{ {\Gamma _\alpha D\gamma _5 } \right\} = \mbox{Sp}\left\{ 
{\frac{1}{2}\left( {n_\alpha - n_\alpha ^ + } \right)D\gamma _5 } \right\}.
\end{equation}

The quantities $D,\;\gamma _5 ,\;\gamma ^\mu $ in relationships (\ref{eq85})-(\ref{eq87}) 
mean $D = Z^ + \tilde {D}Z$, $\gamma _5 = Z^{ - 1}\tilde {\gamma }_5 Z$, 
$\gamma ^\mu = Z^{ - 1}\tilde {\gamma }^\mu Z$. 

In the general case, the matrices $\Gamma _\alpha $ can be expanded over the 
whole set of antisymmetric real matrices related with the DM $\gamma ^\mu = 
Z^{ - 1}\tilde {\gamma }^\mu Z$.

\begin{equation}
\label{eq88}
\Gamma _\alpha = U_\alpha (x) \cdot D^{ - 1} + V_\alpha (x) \cdot \gamma _5 
D^{ - 1} + W_\alpha ^\beta (x) \cdot \gamma _5 \gamma _\beta D^{ - 1}.
\end{equation}

The expansion coefficients $U_\alpha (x),\;V_\alpha (x),\;W_\alpha ^\beta 
(x)$ in (\ref{eq88}) are expressed as

\begin{equation}
\label{eq89}
\left. {\begin{array}{l}
 U_\alpha (x) = \frac{1}{4}\;\mbox{Sp}\left\{ {\Gamma _\alpha D} \right\} \\ 
 V_\alpha (x) = \frac{1}{4}\;\mbox{Sp}\left\{ {\Gamma _\alpha D\gamma _5 } 
\right\} \\ 
 W_\alpha ^\beta (x) = \frac{1}{4}\;\mbox{Sp}\left\{ {\Gamma _\alpha D\gamma 
_5 \gamma ^\beta } \right\} \\ 
 \end{array}} \right\} .
\end{equation}

Comparing relationships (\ref{eq89}) with (\ref{eq85})-(\ref{eq87}) shows that relationships 
(\ref{eq85})-(\ref{eq87}) define all the expansion coefficients of matrix connectivity , 
i.e. coefficients $U_\alpha (x),\;V_\alpha (x),\;W_\alpha ^\beta (x)$.

\begin{equation}
\label{eq90}
\left. {\begin{array}{l}
 U_\alpha (x) = \frac{1}{8}\;\mbox{Sp}\left\{ {\left( {n_\alpha - n_\alpha ^ 
+ } \right)D} \right\} \\ 
 V_\alpha (x) = \frac{1}{8}\;\mbox{Sp}\left\{ {\left( {n_\alpha - n_\alpha ^ 
+ } \right)D\gamma _5 } \right\} \\ 
 W_\alpha ^\beta (x) = \frac{1}{8}\;\mbox{Sp}\left\{ {\left( {n_\alpha - 
n_\alpha ^ + } \right)D\gamma _5 \gamma ^\beta } \right\} \\ 
 \end{array}} \right\} .
\end{equation}

Thus, if we know $Z$ and matrix (\ref{eq84}) at some time, we can find the field of 
matrix connectivity $\Gamma _\alpha $ using formulas (\ref{eq90}).

Thus, we have proven that the quantities $j_\alpha ,\;h_{\alpha \beta } $ 
constructed based on formulas (\ref{eq18}), (\ref{eq19}) from the bispinor matrix $Z$ 
governed by Dirac equation (\ref{eq3}) can be identified with the quantities 
$J_\alpha ,\;H_{\alpha \beta } $, which define the solution of the CGD 
equations provided that the following conditions are fulfilled:

The dimensionless constants $\kappa $ and $\theta $ are related as (\ref{eq79});

The mass $m$ is related to the constant $\lambda $ as (\ref{eq80});

The matrix connectivity $\Gamma _\alpha $ is defined by (\ref{eq88}), (\ref{eq90}).

In conclusion of this section, for convenience of reference, let us list the 
expressions, which relate the vector $J_\alpha $ and the tensor $H_{\alpha 
\beta } $ to the initial quantities $A_\alpha $, $F_{\alpha \beta } $, 
$\lambda $ and to each other, and which result from taking account of (\ref{eq79}), 
(\ref{eq80}), (\ref{eq88}), (\ref{eq90}).

\begin{equation}
\label{eq91}
J_\alpha = \frac{2\theta }{m}A_\alpha ;\quad H_{\alpha \beta } = 
\frac{\theta }{2m^2}F_{\alpha \beta } ,
\end{equation}

\begin{equation}
\label{eq92}
m = \mbox{Const};\quad J^\alpha _{;\alpha } = 0,
\end{equation}

\begin{equation}
\label{eq93}
\left( {J_{\beta ;\alpha } - J_{\alpha ;\beta } } \right) = 4m \cdot 
H_{\alpha \beta } ,
\end{equation}

\begin{equation}
\label{eq94}
H_{\alpha \cdot } {}^\beta _{;\beta } = - m \cdot J_\alpha ,
\end{equation}

\begin{equation}
\label{eq95}
E^{\alpha \beta \mu \nu }H_{\beta \mu ,\nu } \equiv 0.
\end{equation}

\subsection{Cauchy Problem}
\label{subsec:cauchy}

If conditions i-\ref{subsec:cauchy} are 
fulfilled, the statement of the Cauchy problem for the whole set of fields 
differs from the problem statement used in the standard version of the 
quantum field theory (QFT). The major difference lies in the way of finding 
$\Gamma _\alpha $. In QFT, this quantity is found from the Yang-Mills 
equations, in which the sources are current vectors. In our case, the matrix 
connectivity $\Gamma _\alpha $ is derived from relationships (\ref{eq90}). This is 
related to the following:

If conditions i-\ref{subsec:cauchy} are 
fulfilled, equations for the quantities $j_\alpha ,\; h_{\alpha \beta } $ at each time have an autonomous form. Moreover, these equations allow for 
the correct statement of the Cauchy problem for the quantities $j_\alpha 
,\;h_{\alpha \beta } $. 

Using the quantities $j_\alpha ,\;h_{\alpha \beta } $ obtained, the 
Hermitian matrix $M$ (see (\ref{eq20})) is set up, following which the matrix $Z_{ 
}$is found as a solution of algebraic equation (\ref{eq21}). 

Using the explicit form of $Z$, we can calculate its derivatives and 
substitute these into relationship (\ref{eq90}). These particular relationships 
yield expansion coefficients of the matrix connectivity $\Gamma _\alpha $ 
for the full set of anti-symmetric matrices (\ref{eq88}).

The matrix connectivity $\Gamma _\alpha $ calculated using formulas (\ref{eq88}) 
possesses all the attributes of a gauge field corresponding to the group 
$SO(\ref{eq4})$. Indeed, it is easy to demonstrate that gauge transformations

\begin{equation}
\label{eq96}
Z \to {Z}' = ZU^{ - 1},\quad Z^ + \to {Z}'^ + = UZ^ + 
\end{equation}

($U$ are real unitary matrices) transform the matrix connectivity $\Gamma 
_\alpha $ in accordance with the law

\begin{equation}
\label{eq97}
\Gamma _\alpha \to {\Gamma }'_\alpha = U\Gamma _\alpha U^{ - 1} + UU^{ - 
1}_{;\alpha } ,
\end{equation}

\noindent
which is standard for gauge fields. 

\section{Discussion of Results}
\label{sec:discussion}

This paper proposes a solution to one of the matters long discussed by 
physicists and mathematicians -- the matter of physical interpretation of 
the Weyl degrees of freedom of space. Different attempts have been made in 
this area, but no significant progress has been achieved so far. One of the 
approaches provided for relating the Weyl degrees of freedom either to the 
parameters of dark matter and energy in the Universe, or to the cosmological 
red shift, or with scaling of spacetime slice measurements 
(\cite{Sing} - \cite{Fair}). Another 
approach treated new degrees of freedom as an attribute of the integrated 
Weyl space (i.e. space, in which the Weyl vector is a scalar function 
gradient) leading to the appearance of the Shroedinger equation 
(\cite{Santa}, \cite{Novello}). 

The interpretation of the Weyl degrees of freedom proposed in this paper is 
fundamentally new. We propose interpreting these degrees of freedom as 
those, which reproduce the polarization density matrix $M$ of the whole set 
of half-integer spin fields obeying the Dirac equation. The coincidence of 
the CGD equations with the Dirac equation corollaries, which is discussed in 
this paper, in our opinion, is a strong proof of viability of the proposed 
interpretation.

Dynamics of the Weyl degrees of freedom is described by equations (\ref{eq70})-(\ref{eq76}) 
for any structure of the polarization matrix $M$. However, the quantum-field 
interpretation of solutions of these equations has limited applicability; it 
can be used only in the cases when all the eigenvalues of the matrix $M$ are 
positive. This condition corresponds to usual requirements applied to 
polarization density matrices in quantum mechanics and quantum field theory.

The coincidence of the CGD equations with the Dirac equation corollaries 
allows us to draw a conclusion that bispinor states in the quantum field 
theory are just a different language to describe dynamics of the Weyl 
degrees of freedom in some regions of space, where all the eigenvalues of 
the matrix $M$ are positive. When we solve the CGD equations in these 
regions, we in fact solve a quantum field problem, in which half-integer 
spin particles reproduce two quantities of the CGD equations: the vector 
$J_\alpha $ and the anti-symmetric tensor $H_{\alpha \beta } $. 

In conclusion, let us express our vision of where our results are among 
different attempts to unify physical interactions. In our opinion, the 
outcome of this study can be used in theoretical validation of the Standard 
Model (SM) of elementary particles and, in particular, the confinement 
model. Indeed, production and annihilation operators for particles with spin 
$\raise.5ex\hbox{$\scriptstyle 1$}\kern-.1em/ 
\kern-.15em\lower.25ex\hbox{$\scriptstyle 2$} $ and different sets of other 
quantum numbers are obtained by multiplying the bispinor matrix on the right 
by projection operators. The set of projection operators includes $P_\pm = 
\textstyle{1 \over 2}\left[ {E\pm I} \right]$, where $I$ is the reflection 
matrix in the group of invariant transformations $O\left( 4 \right)$ - see 
\cite{Gorb2007}. The projection operators $P_\pm $ have ranks 
1 and 3, and allow splitting four bispinors $ZP_{\eta ,\lambda } \quad 
\left( {\eta ,\lambda = \pm } \right)$ into two groups. One group comprises 
three states that differ in two quantum numbers. The second group includes 
one state with zero quantum numbers (the "sterile" state similar to the 
right neutrino). Such splitting of states coincides with splitting in each 
generation of leptons and quarks within the SM. In the context of 
understanding the confinement phenomenon, it is important that bispinor and 
matrix connectivities exist only in the space regions, where the 
polarization density matrix is positively defined. Apparently, different 
states in such space-localized fields of the bispinor matrix and matrix 
connectivity can be correlated with different elementary particles.

\section{Acknowledgements}

The author thanks his colleagues A.A. Sadovoy and A.K. Khlebnikov for a 
number of helpful discussions of the issues considered in this paper.

\end{document}